\documentclass[aps,pra,twocolumn,superscriptaddress,showpacs]{revtex4}
\usepackage{tipa}
\usepackage{pifont}
\usepackage{txfonts}
\usepackage{amssymb}
\usepackage{graphicx}

\begin{document}

\title{Partial-state fidelity and quantum phase transitions
induced by continuous level crossing}

\author{Ho-Man Kwok}
\affiliation{Department of Physics and ITP, The Chinese University of Hong
Kong, Shatin, Hong Kong, China}

\author{Chun-Sing Ho}
\affiliation{Department of Physics and ITP, The Chinese University of Hong
Kong, Shatin, Hong Kong, China}

\author{Shi-Jian Gu}
 \email{sjgu@phy.cuhk.edu.hk}
\affiliation{Department of Physics and ITP, The Chinese University of Hong
Kong, Shatin, Hong Kong, China}

\date{\today}

\begin{abstract}
The global-state fidelity cannot characterize those quantum phase
transitions (QPTs) induced by continuous level crossing due to its
collapse around each crossing point. In this paper, we take the
isotropic Lipkin-Meshkov-Glick (LMG) model and the
antiferromagnetic one-dimensional Heisenberg model as examples to
show that the partial-state fidelity can signal such
level-crossing QPTs. Extending to the thermodynamic limit we
introduce the partial-state fidelity susceptibility and study its
scaling behavior. The maximum of the partial-state fidelity
susceptibility goes like $N$ for the LMG model and $N^3$ for the
Heisenberg model.
\end{abstract}

\pacs{64.60.-i, 05.70.Fh, 75.10.-b}




\maketitle

It has been an interesting issue for merging quantum phase transitions (QPTs)
and fidelity. The former one is noticed by the observation of quantities
undergoing structural changes around some critical points \cite{sachdev}, while
the latter one measures the amount of relevance for two quantum ground states
of the system differed by some parameters \cite{nilesen}. A physical phenomenon
is then associated with a pure quantum informational consideration
\cite{HTQuan06,PZanardi06,HQZhou07,HQZhou071}. The fidelity approach brings
advantages to the characterization of QPTs, because by comparing the states, no
\textit{a priori} knowledge to the order parameter, symmetry, and type of QPTs
of the system are required. The fidelity approach has been examined in various
models and proved its ability in characterizing QPTs \cite{PZanardiStat,
Buonsante07, MCozzini07, MCozzini07b, Chen07}. This suggests experimental
measurements of the quantum state itself which is a rather challenging task.
Leading order of the fidelity has been suggested as well \cite{PZanardi07FS,
WLYou07FS}, for its critical exponents and divergence helps classification of
the universality of the system \cite{LCVenuti07FS, SJGu07FS, MFYang07FS,
MFYang08FS, HMKwok08FS}.

However, there are still limitations to the fidelity approach. It is useful to
study ground states of some continuous variables, but unable to describe
discrete global ground states, i.e. states with fixed quantum numbers within a
certain continuous range of parameters. Especially when the driving Hamiltonian
commutes with the whole Hamiltonian, the leading order of the fidelity is not
well-defined. To tackle this problem, the partial-state fidelity was introduced
\cite{HQZhou071,NPaunkovic08}. It concerns the quantum relevance of part of a
system with respect to a global change of the parameter. It has been
investigated in characterizing the QPTs in the XY model, a three-body
interacting model \cite{HQZhou071} and a conventional BCS superconductor with
an inserted magnetic impurity system \cite{NPaunkovic08}. In addition, the
operator fidelity susceptibility was also introduced and was shown that it can
signal QPTs regardless of the degeneracy of the system \cite{XWang08}. However,
attention to those phase transitions of continuous level crossing were not paid
under their definitions.

In this paper, we put our attention on thermodynamic systems in
which continuous level crossing occurs. One is the isotropic
Lipkin-Meshkov-Glick (LMG) model introduced in nuclear physics
\cite{LMG}, which is related to Bose-Einstein condensation and
Josephson junctions. We make use of its exact spectrum to obtain
the partial-state fidelity. The other one is the one-dimensional
Heisenberg model, where we adopt the Bethe-Ansatz method to
compute the ground state energy, as well as the required reduced
density matrix. We show that the partial-state fidelity can be
used to locate the critical point for these two models. We
defined the corresponding fidelity susceptibility and perform
scaling analysis.

Let a system be parameterized by $h$, with its density operator
$\hat{\rho}(h) = \left|\Psi (h) \right\rangle \left\langle \Psi
(h)\right|$ corresponds to the ground state $\left|\Psi
(h)\right\rangle$. When $h$ is displaced by $\delta h$ such that
$\tilde{h} = h+\delta h$, the density operator becomes
$\hat{\rho}(\tilde{h}) = \left|\Psi (\tilde{h})\right\rangle
\left\langle \Psi(\tilde{h})\right|$, the fidelity is defined
according to their respective density operator
\begin{eqnarray}
F(h,\tilde{h}) &=&
\textrm{Tr}\sqrt{\sqrt{\hat{\rho}(h)}\hat{\rho}(\tilde{h})
\sqrt{\hat{\rho}(h)}} \nonumber \\
&=& |\langle\Psi(h)|\Psi(\tilde{h})\rangle|.
\end{eqnarray}
If the system is divided into two subsystems $A$ and $B$, the
reduced density operator $\hat{\rho}_A(h) = \textrm{Tr}_B
\hat{\rho}(h)$ contributes to the partial-state fidelity
\begin{eqnarray}
F_A\left(h,\tilde{h}\right) =
\textrm{Tr}\sqrt{\sqrt{\hat{\rho}_A(h)}\hat{\rho}_A(\tilde{h})
\sqrt{\hat{\rho}_A(h)}}. \label{eq:PSFdef}
\end{eqnarray}

The partial state of subsystem $A$ can be a single-site state or a
two-site state, or even a larger subsystem state. For convenience
in this paper we consider tracing out all particles but one. So
for a system with definite magnetization $M$, one can make use of
the on-site average magnetization basis $\langle \sigma^z \rangle
= 2M/N$, where $N$ is the number of spins, to trace out the
density operator. This left us the diagonal reduced density
matrix $\rho_A(h)$
\begin{eqnarray}
\rho_A(h) = \frac{1}{2}\left( \begin{array}{cc}
1+\langle\sigma^z\rangle & 0
\\ 0 & 1 - \langle\sigma^z\rangle
\end{array} \right). \label{eq:RDMsigma}
\end{eqnarray}

Consider a system of size $N$ with a set of discrete ground state
level-crossing points $\{h_j\}$, where $j = 0,1,2...$ and $h_j >
h_{j+1}$. Let the partial state within a range $h \in R_j^{(N)} =
(h_{j}, h_{j-1})$ with an average magnetization $\langle \sigma_z
\rangle^j$, the partial-state fidelity at $h_j$ is defined by

\newpage
\begin{eqnarray}
&&F_A (h_j) = \textrm{Tr}\nonumber \\
&&\sqrt{\frac{1}{4}\left( \begin{array}{cc}
1+\langle\sigma^z\rangle^j & 0
\\ 0 & 1 - \langle\sigma^z\rangle^j
\end{array} \right)\left( \begin{array}{cc}
1+\langle\sigma^z\rangle^{j+1} & 0
\\ 0 & 1 - \langle\sigma^z\rangle^{j+1}
\end{array} \right)} \nonumber \\
&=& \frac{1}{2}\sqrt{\left(1+\langle \sigma_z \rangle^j\right)\left(1+\langle
\sigma_z \rangle^{j+1}\right)}
 \nonumber \\
&& + \frac{1}{2}\sqrt{\left(1-\langle \sigma_z
\rangle^{j}\right)\left(1-\langle \sigma_z \rangle^{j+1}\right)}.
\label{eq:PSF}
\end{eqnarray}
It is the trace of the reduced density matrices at two sides. The
non-unity of Eq. (\ref{eq:PSF}) signals the level crossing when
$\langle\sigma^z\rangle$ changes.

{\it The isotropic LMG model:} The model reads
\begin{eqnarray}
 H_{\rm LMG} &=&  - \frac{1}{N}\sum\limits_{i < j} {\left( {\sigma _x^i \sigma _x^j
 + \sigma _y^i \sigma _y^j } \right)}  - h\sum\limits_i {\sigma _z^i}
 \nonumber  \\
 &=&  - \frac{2}{N}\left( {S_x^2  + S_y^2 } \right) - 2hS_z  + \frac{1}{2}
 \nonumber  \\
 &=&  - \frac{2}{N}\left( {\textbf{S}^2  - S_z^2  - N/2} \right) -
 2hS_z. \label{eq:LMGmodel}
\end{eqnarray}
For $\sigma _{\kappa}\, (\kappa=x,y,z)$ are the Pauli matrices,
and $S_\kappa = \sum _i \sigma _{\alpha}^i/2$. It describes a
system of mutually interacting spins subjected to a transverse
external field of strength $h$. The ground state lies in the
maximum spin sector $S = N/2$. For it is diagonal in the basis
$\left|N/2,M\rangle\right.$, its eigenenergies are given by
\begin{eqnarray}
E\left( {M,h} \right) = \frac{2}{N}\left( {M - \frac{hN}{2}}
\right)^2 - \frac{N}{2}\left( {1 + h^2 } \right).
\end{eqnarray}
The ground state is determined by the minimum of the square,
\begin{eqnarray}
M_0  = \left\{ \begin{array}{cc}
 \Large{\mbox{$\frac{N}{2}$}} & h \ge 1 \\
 \\
 I\Large{\mbox{$\left( \frac{hN}{2} \right)$}} & 0 \le h < 1 \\
 \end{array} \right.,
\end{eqnarray}
where $I(x)$ gives the integer part of $x$. It can be shown that
ground state level crossing occurs at some $h_j \equiv 1-(2j+1)/N$
and when $h \in (h_j, h_{j-1})$, the ground state is $M = N/2 -
j$ \cite{HMKwok08FS, JVidal05}. In the thermodynamic limit $h =
1$ is the critical point. Since the model [Eq.
(\ref{eq:LMGmodel})] is infinitely coordinated, we consider the
partial state as a single particle state with density matrix still
follows Eq. (\ref{eq:RDMsigma}). The partial-state fidelity at
$h_j$ has the exact form according to Eq. (\ref{eq:PSF})
\begin{eqnarray}
F_A = \frac{1}{N}\left(\sqrt{(N-j)(N-j-1)}+\sqrt{j(j+1)}\right).
\end{eqnarray}

\begin{figure}
\includegraphics[width=\linewidth]{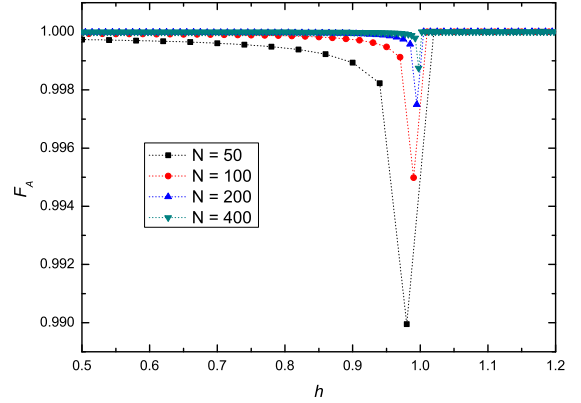}
\caption{Partial-state fidelity as a function of external field
strength $h$ of the LMG model with different system sizes.}
\label{fig:FLMG}
\end{figure}

Fig. \ref{fig:FLMG} shows the plot based on the above formula.
$F_A$ drops to a minimum at $h = h_0$, the level-crossing point
closest to the critical point $h_c = 1$. Since there are no
further level-crossing points for $h > 1$, the partial-state
fidelity maintains the value one. As system size increases, the
minimum of $F_A$ gets closer to one. It is because unlike
ordinary treatment in the fidelity where $\delta h$ is fixed, we
calculate the partial-state fidelity obtained by two nearest
level-crossing points. When $\delta h$ becomes smaller, the
similarity between states becomes higher and is reflected by the
close-to-unity behavior, similar to that in global fidelity.

We emphasis the comparison between neighboring partial states, as
a realization to continuous level crossing when $N \to \infty$.
The partial-state fidelity helps extrapolating discrete level
crossing to continuous level crossing.

\begin{figure}
\includegraphics[width=\linewidth]{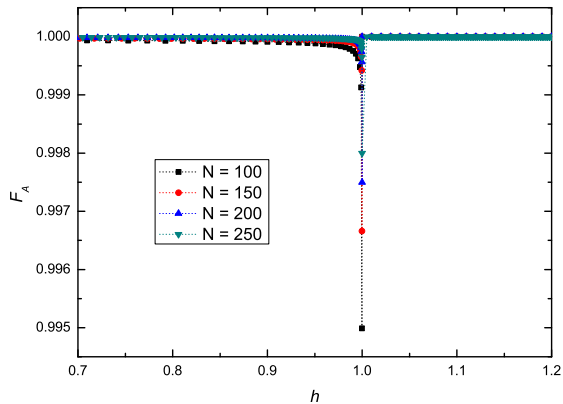}
\caption{Partial-state fidelity as a function of external field
strength $h$ of the 1D Heisenberg model.} \label{fig:FHei}
\end{figure}

{\it The one dimensional Heisenberg model:} The isotropic LMG
model provides us an analytic form of the partial-state fidelity.
Next we try to examine the one dimensional Heisenberg model,
which is another system that exhibits ground state level crossing.
The Hamiltonian reads
\begin{eqnarray}
H_{\rm Heisenberg} &=& \sum_i^N S_i^x S_{i+1}^x + S_i^y S_{i+1}^y
+ S_i^z S_{i+1}^z - 2hS_i^z,
\end{eqnarray}
where $S_i^\kappa$ is the spin-1/2 operator at site $i$. With a ring geometry
$S_{N+1}^\kappa = S_1^\kappa$ imposed, one can solve the spectrum by the
Bethe-Ansatz method \cite{HABethe31}
\begin{eqnarray}
E_0=\frac{N}{4}-Nh+\sum_{j=1}^{N_\downarrow}
\left(2h-\frac{2}{x_j^2 +1}\right) \label{eq:BAGE}
\end{eqnarray}
where $N_\downarrow$ is the number of down spins, and $x_j$ are
spin rapidities. They satisfy the Bethe ansatz equations
\cite{HABethe31}
\begin{eqnarray}
2N\tan^{-1} x_j=2\pi I_j
+2\sum_{l=1}^{N_\downarrow}\tan^{-1}\frac{x_j-x_l}{2},
\label{eq:rapidities}
\end{eqnarray}
where $I_j$ are quantum numbers and take values of
$-(N_\downarrow -1)/2, \cdots, (N_\downarrow -1)/2$ for the
ground state. By solving the Bethe-Ansatz equations numerically,
the eigenenergies are obtained and the ground state is determined
by the minimum eigenenergy. So are the level-crossing points. For
the single-site subsystem $A$ can also be characterized by the
on-site magnetization $\langle \sigma^z \rangle$, according to
Eq. (\ref{eq:PSF}) again we compute the partial-state fidelity at
the level-crossing points by comparing the nearest partial states.
The numerical result is shown in Fig. \ref{fig:FHei}. We find
similar features as in the LMG model such as a drop of $F_A$ at
the critical point, the drop becomes sharper and the minimum gets
closer to one as system size increases.

{\it Partial-state fidelity susceptibility:} Interested in the
continuous level crossing that corresponds to the thermodynamic
limit, we introduce the concept of partial-state fidelity
susceptibility. It is because in such case an infinitesimal
change of the parameter is sufficiently responsible for an
obvious change of the fidelity. The partial-state fidelity
susceptibility is defined in a similar manner as \cite{WLYou07FS}:
\begin{eqnarray}
\chi_{_F}^{(A)} = \lim_{\delta h \to 0}
\frac{-2\textrm{ln}F_A}{(\delta h)^2}. \label{eq:PSFS}
\end{eqnarray}
The above formula combines the ability of partial-state fidelity
in observing level-crossing transitions and the idea of
global-state fidelity susceptibility that measures the leading
response of fidelity to infinitesimal change of parameter.

In finite systems, we compute $\chi_{_F}^{(A)}$ by taking the
natural log of the partial-state fidelity at $h_j$, and the divide
it by the square of the modulus of the range $R_{j+1}^N$, i.e.
$\delta h = h_j - h_{j+1}$. With this notion we arrive the
analytic form of $\chi_{_F}^{(A)}$ of the LMG model as $\delta h
= 2/N$:
\begin{eqnarray}
\chi_{_F}^{(A)} =
-\frac{N^2}{2}\textrm{ln}\left[\sqrt{\left(1-\frac{j}{N}\right)
\left(1-\frac{j+1}{N}\right)}+\sqrt{\frac{j(j+1)}{N^2}}\right].
\label{eq:FSLMG}
\end{eqnarray}
\begin{figure}
\includegraphics[width=\linewidth]{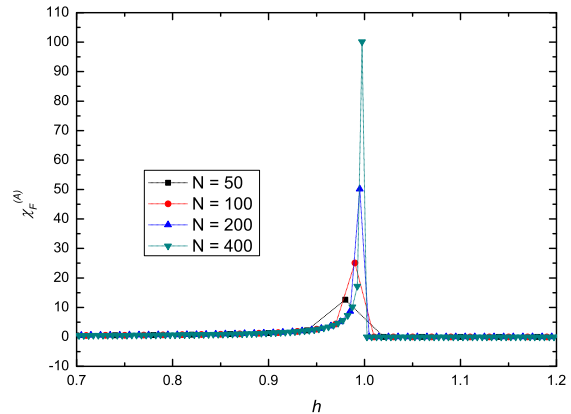}
\caption{Partial-state fidelity susceptibility as a function of
external field strength $h$ of the LMG model with different
system sizes.} \label{fig:FSLMG}
\end{figure}

The plot in fig. \ref{fig:FSLMG} shows $\chi_{_F}^{(A)}$ grows
with system size, and arrives its maximum at $h_0$, the
level-crossing point closest to the critical point. The response
near the maximum becomes sharper for larger systems, indicating a
divergence in the thermodynamic limit. It suggests
$\chi_{_F}^{(A)}$ as a smooth function of $h$ in the
thermodynamic limit except at the critical point. It diverges at
$h = h_c$ and drops to zero when $h > h_c$. The divergence of the
maximum goes like $N$, since at $j = 0$, from Eq. (\ref{eq:FSLMG})
\begin{eqnarray}
-\frac{N^2}{2}\textrm{ln}\sqrt{1-\frac{1}{N}} = \frac{N}{4}
\end{eqnarray}
for large $N$.

\begin{figure}
\includegraphics[width=\linewidth]{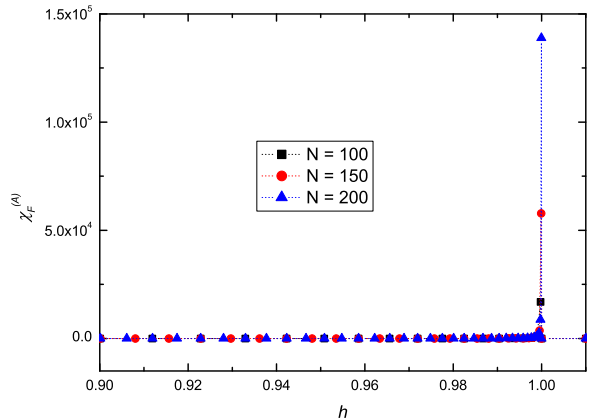}
\caption{Partial-state fidelity susceptibility as a function of
external field strength $h$ of the 1D Heisenberg
model.}\label{fig:FSHei}
\end{figure}

We compute $\chi_{_F}^{(A)}$ for the Heisenberg model and the
result is shown in Fig. \ref{fig:FSHei}. The divergence is even
sharper. Although the full analytic form of the $\chi_{_F}^{(A)}$
is inaccessible as the spin rapidities $x_j$ form a set of
transcendental equations, the critical exponent can be estimated
by obtaining $h_0 - h_1$.

For $N_\downarrow = 1$, from Eq. (\ref{eq:rapidities}), we have
\begin{eqnarray}
2N\tan^{-1} x_1 = 0, \;\,\,\,\, x_1 = 0.
\end{eqnarray}
The ground state energy for $N_\downarrow = 0$ is simply
$\frac{N}{4} - Nh$ and that of $N_\downarrow = 1$ is calculated by
Eq. (\ref{eq:BAGE})
\begin{eqnarray}
\frac{N}{4} - Nh + 2(h-1), \label{eq:GENdown1}
\end{eqnarray}
in which $h_0 = 1$ is determined.

For $N_\downarrow = 2$, Eq. (\ref{eq:rapidities}) consists of two
equations
\begin{eqnarray}
2N\tan^{-1}x_1 &=& -\pi + 2\tan^{-1}\left(\frac{x_1 -x_2}{2}\right)
\nonumber \\
2N\tan^{-1}x_2 &=& \pi + 2\tan^{-1}\left(\frac{x_2 -
x_1}{2}\right).
\end{eqnarray}
The value of $x_1$ and $x_2$ can be found, since by symmetry $x_1
= -x_2$, the above two equations become one
\begin{eqnarray}
2N\tan^{-1}x_2 = \pi + 2\tan^{-1}x_2, \;\,\,\,\, x_2 =
\tan\frac{\pi}{2(N-1)}.
\end{eqnarray}
The ground state energy for $N_\downarrow = 2$ is
\begin{eqnarray}
\frac{N}{4} - Nh + 2\left(2h -
\frac{2}{\left[\tan\frac{\pi}{2(N-1)}\right]^2+1} \right).
\label{eq:GENdown2}
\end{eqnarray}
Then $h_1$ is determined when Eq. (\ref{eq:GENdown1}) equals to
Eq. (\ref{eq:GENdown2}), that is
\begin{eqnarray}
h_1 = -1 + \frac{2}{\left[\tan\frac{\pi}{2(N-1)}\right]^2+1}.
\end{eqnarray}
Expanding for large $N$ limit, for $\tan y \simeq y$ for small
$y$, we have $h_1 \simeq 1 - \frac{\pi^2}{2(N-1)^2}$ and thus
\begin{eqnarray}
\delta h = h_0 - h_1 \simeq \frac{\pi^2}{2(N-1)^2}.
\end{eqnarray}
The partial-state fidelity susceptibility scales like
\begin{eqnarray}
\chi_{_F}^{(A)} =
-\frac{8(N-1)^4}{\pi^4}\textrm{ln}\sqrt{1-\frac{1}{N}} \propto
N^3,
\end{eqnarray}
which is apparently different from that of the LMG model.

Let us make a remark. In many times, it is often to consider some
averaged physical quantities to understand the intrinsic response
to the driving agent. Fidelity susceptibility is one of them
\cite{LCVenuti07FS, SJGu07FS}. But for the partial-state fidelity
susceptibility, we have already focused on a part of the system.
Such a local response to the global driving has already played a
role as a certain averaged quantity. So we believe, supported by
the two distinct models above, the divergence of the partial-state
fidelity in continuous level crossing could be a general feature.
Its divergence in the isotropic LMG model may be related to the
$\gamma \to 1$ limit of the averaged fidelity susceptibility
driven by external field in \cite{HMKwok08FS}.

We introduced the partial-state fidelity susceptibility formalism
derived from the partial-state fidelity and showed it is a
suitable candidate to describe quantum phase transitions induced
by continuous level crossing, which global-state fidelity cannot
provide information to. Focusing on a subsystem, the
partial-state fidelity susceptibility still diverges in the
thermodynamic limit. The sudden drop-to-zero indicates the
critical point of the system.

By examining two models, the isotropic LMG model and the one
dimensional Heisenberg model, we started from the discrete level
crossing and extrapolated to the thermodynamic limit which
corresponds to continuous level crossing. We find the maximum of
the partial-state fidelity susceptibility goes like $N$ for the
LMG model and $N^3$ for the Heisenberg model, indicating they
belong to different universality classes. The former one could be
treated as a complement to the fidelity susceptibility analysis
in the LMG model \cite{HMKwok08FS}.

We demonstrated the calculation for a single-site partial state.
However, the partial-state fidelity susceptibility shall not be
limited to (sub)systems with definite magnetization, because it
can still be well-defined for two-particle or many-particle
partial states, yet not for all-particle (global) states. The
partial-state fidelity is a new approach to tackle QPTs, studying
its leading order which is independent of the small change of the
driving parameter helps understanding the continuous level
crossing QPTs as well as to determine the critical points. We
hope this encourages discussions on the related topics.

{\it Note added:} After finishing this work, we received a preprint from XG
Wang, in which the fidelity and its susceptibility of two-site partial state in
the LMG model are studied \cite{XGWang}.

We thank J. Vidal for comments on our work. This work is supported
by the Direct grant of CUHK (A/C 2060344)

\end{document}